\documentclass[a4paper]{cas-dc}

\usepackage[numbers]{natbib}

\usepackage{graphicx}
\usepackage{float}
\usepackage{subfigure}
\usepackage{pgfplots}
\usepackage{multirow}
\usepackage{multicol}
\usepackage{enumerate}
\usepackage{booktabs}
\usepackage{setspace}
\usepackage{amsmath}
\usepackage{amsfonts}
\usepackage{amssymb}
\usepackage{makecell}
\usepackage{diagbox}
\usepackage{ulem}

\usepackage{enumitem}
\usepackage{xcolor}
\definecolor{myblue}{RGB}{0,114,189}
\definecolor{myyellow}{RGB}{216,83,25}
\definecolor{myred}{RGB}{237,177,32}
\definecolor{mypurple}{RGB}{128,0,128}

\usepackage{soul}

\def\tsc#1{\csdef{#1}{\textsc{\lowercase{#1}}\xspace}}
\tsc{WGM}
\tsc{QE}

\begin{document}
\let\WriteBookmarks\relax
\def\floatpagepagefraction{1}
\def\textpagefraction{.001}
\let\printorcid\relax    
\shorttitle{KAMCL} 

\shortauthors{}  

\title [mode = title]{Knowledge-Aware Multi-Intent Contrastive Learning for Multi-Behavior Recommendation}  

\tnotemark[1,2]

\tnotetext[1]{This work was supported by the Innovation Capability Improvement Plan Project of Hebei Province (No. 22567637H), the S\&T Program of Hebei(No. 236Z0302G), and HeBei Natural Science Foundation under Grant (No.G2021203010 \& No.F2021203038).}

%

\author[1,2]{Shunpan Liang}[style=chinese]


\ead{liangshunpan@ysu.edu.cn}

\ead[url]{http://web.ysu.edu.cn/LSP/zh_CN/zhym/50462/list/index.htm}

\credit{Conceptualization of this study, Methodology, Software}

\author[1]{Junjie Zhao}[style=chinese]
\ead{zhaojunjie1314@stumail.ysu.edu.cn}

\author[1]{Chen Li}[style=chinese]
\ead{lichen36211@gmail.com}
\cormark[1]

\credit{Data curation, Writing - Original draft preparation}

\author[1]{Yu Lei}[style=chinese]
\ead{leiyu0160@gmail.com}

\credit{Data curation, Writing - Original draft preparation}

\affiliation[1]{organization={School of Information Science and Engineering, Yanshan University},
    city={Qinhuangdao, Hebei Province},
    postcode={066000}, 
    country={China}}

\affiliation[2]{organization={School of Information Science and Engineering, Xinjiang College Of Science \& Technology},
    city={Korla, Xinjiang},
    postcode={841000}, 
    country={China}}

\cortext[cor1]{Corresponding author}


\begin{abstract}
Multi-behavioral recommendation optimizes user experiences by providing users with more accurate choices based on their diverse behaviors, such as view, add to cart, and purchase. Current studies on multi-behavioral recommendation mainly explore the connections and differences between multi-behaviors from an implicit perspective. Specifically, they directly model those relations using black-box neural networks. In fact, users' interactions with items under different behaviors are driven by distinct intents. For instance, when users view products, they tend to pay greater attention to information such as ratings and brands. However, when it comes to the purchasing phase, users become more price-conscious. 
To tackle this challenge and data sparsity problem in the multi-behavioral recommendation, we propose a novel model: Knowledge-Aware Multi-Intent Contrastive Learning (KAMCL) model. This model uses relationships in the knowledge graph to construct intents, aiming to mine the connections between users' multi-behaviors from the perspective of intents to achieve more accurate recommendations.
KAMCL is equipped with two contrastive learning schemes to alleviate the data scarcity problem and further enhance user representations. 
Extensive experiments on three real datasets demonstrate the superiority of our model.

\end{abstract}





\maketitle
 
\section{INTRODUCTION}
Recommender System (RS) plays a vital role in delivering personalized user services and alleviating information overload. The fundamental challenge of RS lies in extracting user preferences from behavioral data to provide accurate recommendations. Traditional approaches rely on collaborative filtering techniques\cite{su2009survey,kipf2016semi,he2020lightgcn,wang2019neural}, which are primarily suitable for modeling single behavior, such as click or purchase. However, in real-life scenarios, users tend to have multiple behaviors. For instance, in the e-commerce domain, users engage with items through various behaviors including view, add to cart, and purchase. Yet, the acquisition of the \textbf{target behaviors} (i.e., purchase) could be expensive, consequently leading to data sparsity problems\cite{CML2022}. In contrast, capturing \textbf{auxiliary behaviors} (i.e., view, add to cart) is cost-effective and could supplement valuable user preference information. Therefore leveraging the abundant auxiliary behaviors to uncover the relationship between these behaviors and target behaviors becomes paramount in generating more accurate user preferences for multi-behavior recommendations.

Recently, numerous research efforts have emerged in multi-behavior recommendations. 
Some studies employ graph construction to unveil latent relationships between diverse behaviors. 
For instance, MBGCN~\cite{MBGCN2020} constructs a unified graph to capture auxiliary behaviors’ different influence strengths on target behavior. 
Some other works utilize graph neural networks to explore heterogeneous relationships between different behaviors~\cite{MBGCN2020,MBGMN2021,GNMR2021,yan2023cascading}.
Others employ Knowledge graphs to enhance the item attribute information used in multi-behavior recommendations~\cite{KHGT2021,KMCLR2023,CKML2023}. 
Besides, transformer networks~\cite{vaswani2017attention} and contrastive learning techniques have been employed to enhance the performance of multi-behavior recommendations~\cite{MATN2020,CML2022,MMCLR2022}.



While the outlined approaches above are effective, two challenges still persist.
\begin{figure}[t]
\centering
\includegraphics[width=1\linewidth]{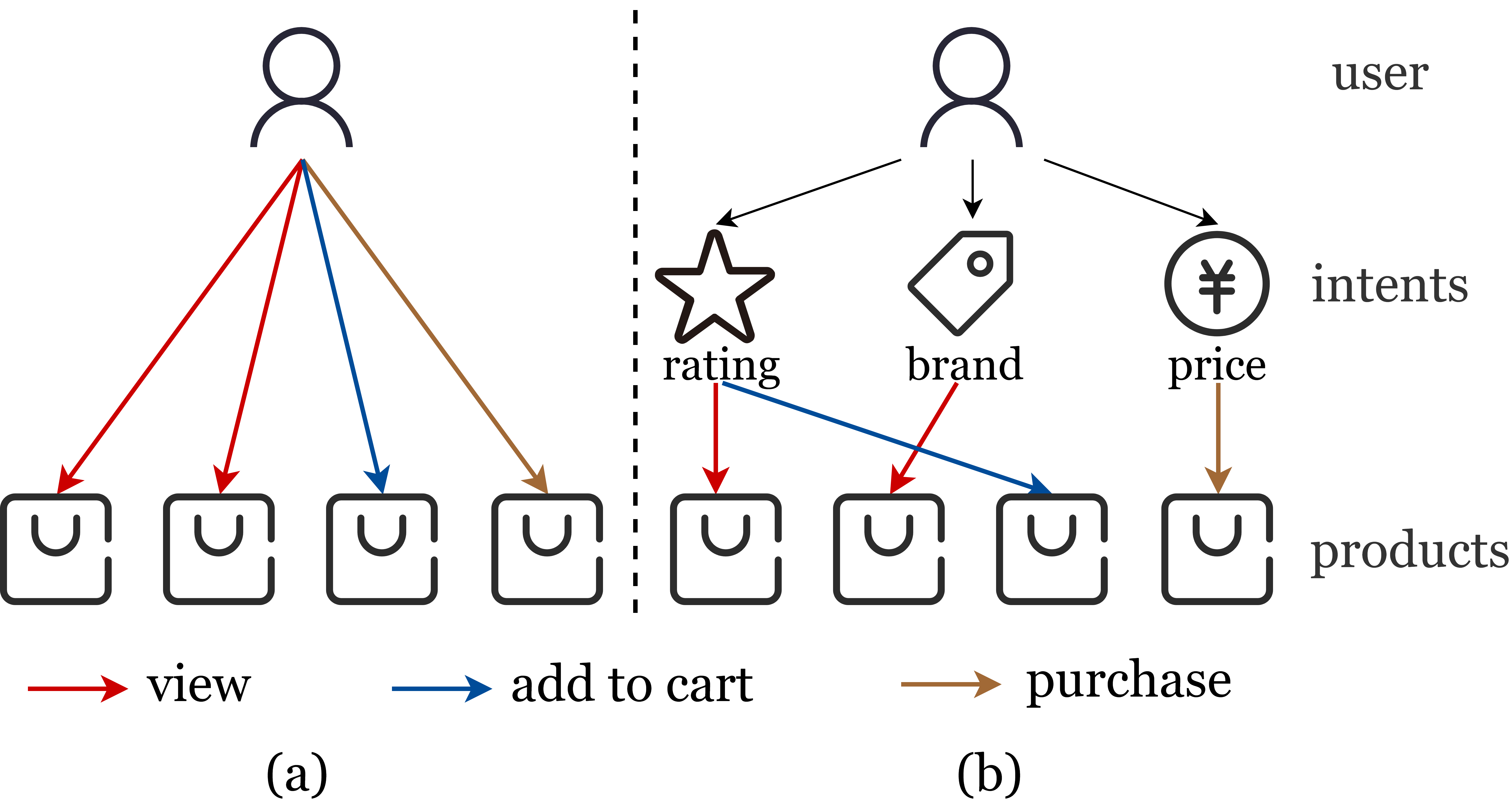}
\vspace{-0.1in}
\caption{Illustrations of the difference between the conventional approaches and our approach for multi-behavior recommendation. (a) depicts the commonly used manner, and (b) depicts our method which considers the intents of the user when making different behaviors.}\label{fig:picture}
\label{figure}
\vspace{-0.1in}
\end{figure}

(1) Intents of multiple types of behaviors are often overlooked. The majority of existing methods primarily focus on examining the connections and differences between behaviors from an implicit perspective. For instance, some approaches\cite{MBGCN2020,GNMR2021,MBRec2022} directly utilize graph neural networks to uncover dependencies between different behaviors in a user-item interaction graph, as shown in Fig.~\ref{fig:picture} (a). However, it is crucial to recognize that users' interactions with items under different behaviors are driven by distinct intents. As depicted in Fig.~\ref{fig:picture} (b), when users view products (red line), they tend to pay greater attention to information such as rating and brand. However, when it comes to the purchasing phase (brown line), users become more price-conscious. Unfortunately, the exploration of correlations between behaviors from the intent perspective is often overlooked in current approaches.

(2) The data sparsity problem in multi-behavior recommendation is not sufficiently addressed. Currently, there are primarily two approaches to address the sparsity issue in multi-behavior data. The first method involves utilizing knowledge graphs as additional information to tackle the problem of sparse item attributes. However, most existing methods only leverage the knowledge graph to construct item-item relationship graphs~\cite{KHGT2021, CKML2023}, without fully exploring heterogeneous information, such as relations and attributes. Another approach aims to alleviate sparsity between behaviors through contrastive learning~\cite{CML2022,MMCLR2022}. Leveraging the outstanding performance of contrastive learning in the field of self-supervised learning~\cite{bengio2009learning}, this method focuses on mining the heterogeneity between the target and auxiliary behaviors. The effectiveness of this technique relies significantly on choosing a suitable comparison scheme. Well-designed schemes for extracting positive and negative samples showcase enhanced performance in capturing user preferences.

\begin{sloppypar}
To tackle the two challenges mentioned above, we propose \textbf{K}nowledge-\textbf{A}ware \textbf{M}ulti-Intent \textbf{C}ontrastive \textbf{L}earning (KAMCL) for multi-behavior recommendation. Specifically, to explore the connections between behaviors from the intent perspective and fully leverage the knowledge graph, KAMCL incorporates a relation-aware knowledge aggregation module to capture information about the attributes of an item from different relations. By leveraging the relational information in the knowledge graph, an intent generation module is designed to capture user intent information across different behaviors. Finally, the intent-based multi-behavior interaction module is used to find the connections between behaviors from the perspective of intent. Meanwhile, KAMCL also designs two comparative learning schemes to address sparsity in knowledge graphs and interactive data.
\end{sloppypar}

To summarize, this work makes the following contributions:
\begin{itemize}
\item[$\bullet$]We propose a new framework for the multi-behavior recommendation that mines the dependencies between user behaviors from the intent perspective based on knowledge graphs.

\item[$\bullet$]We design two contrastive learning schemes, namely relation-aware item contrastive learning and behavior-aware user contrastive learning, to address the data sparsity problem in the model.

\item[$\bullet$]We conduct extensive experiments on three public datasets to demonstrate that our proposed KAMCL significantly outperforms the state-of-the-art, with average lifts of 12.15\% and 17.19\% on HR and NDCG, respectively.
\end{itemize}

\section{RELATED WORK}
\subsection{Multi-behavior Recommendation}

In people's daily lives, various interactive behaviors are accompanied by implicit user preference information. Many multi-behavioral recommendation methods have been proposed to tap into the underlying user preferences in these behaviors. Some traditional approaches employ collaborative filtering techniques, while with the advancements in deep learning, neural networks, graph neural networks, and self-attention mechanisms are increasingly prevalent in the field of multi-behavior recommendation~\cite{zhou2018atrank,NMTR2019,tang2016empirical,zhang2023denoising}. For example, NMTR\cite{NMTR2019} models multiple behaviors based on neural networks and uses cascading to capture relationships between behaviors. After that, MATN\cite{MATN2020} proposes a memory attention network that contextualizes signals between different types of behavior, while MBGCN\cite{MBGCN2020}, MBRec\cite{MBRec2022} apply graph neural networks (GNN) design models modeling user multi-type behavior.

\begin{sloppypar}
    In recent years, many researchers have attempted to introduce intent concept into the recommendation domain\cite{fan2019metapath, wang2020disenhan, wang2020disentangled}. For instance, ICM-SR~\cite{ICM-SR2020} treats the initial item prediction generated through a session as intent and designs an intent-guided detector to assist in session recommendation. ISRec~\cite{ISRec2022} employs a transformer-based approach to extract the intent of the target user from sequential contexts and obtain the future intentions. MIDGN~\cite{MIDGN2022} disentangles user intents from both global and local perspectives to better represent user preferences for bundle recommendation. But in the field of multi-behavioral recommendation, work in this area is still scarce. DIPN\cite{guo2019buying} uses hierarchical attention mechanism to predict the user's purchase intent. CKML\cite{CKML2023} incorporates multi-interest learning into multi-behavioral recommendation, learning shared and specific interests under different behaviors. Inspired by the above work, we design an intention-level network model that aims to reveal connections among users' intentions under different behaviors and use the combination of relationships in the knowledge graph as a composition of our intents. Unlike existing work, we consider the user's intention under different behaviors and based on that we explore the connection between different behaviors.
   
\end{sloppypar}

\subsection{Knowledge Graph-based Recommendation}
    With their rich item-side information and significant potential for interpretability, knowledge graphs are extensively utilized in various domains of recommender systems. KGAT\cite{KGAT2019} models higher-order connections in networks by constructing knowledge graph attention networks in an end-to-end manner. KHGT\cite{KHGT2021} proposes knowledge-enhanced hierarchical graph transformation networks to research multiple kinds of user-item interactions for multi-behavior recommendation. Additionally, there are methods that integrate knowledge graphs with comparative learning and have achieved relatively good results~\cite{KACL2023,KGCL2022,KMCLR2023}. These works leverage the richness of side information in the knowledge graph to better integrate different aspects of semantic information, thereby generating item representations with more valuable information.
\subsection{Contrastive Learning}
    Contrastive learning aims to distinguish similarities and differences between objects by comparing the differences between positive and negative samples in different views\cite{KMCLR2023}. There are many approaches that use contrast learning to optimize the performance of models\cite{MMCLR2022,CML2022,long2021social,qiu2022contrastive}. As a result, in the field of multi-behavioral recommendation, comparative learning becomes an important tool for optimizing models. For example, MMCLR\cite{MMCLR2022} designs contrast learning methods for sequence views versus graph views so that temporal and structural information can be included in the user/item representation. CML\cite{CML2022} fuses meta-learning with contrastive learning to encode the heterogeneity between different user behaviors. In this work, we design a knowledge graph-based relationship-aware contrast learning scheme to generate better item feature and intent representations for multi-behavior recommendation.
    
\section{PROBLEM DEFINITION}
In this paper, we let $U=\{u_1, u_2, ..., u_N\}$ and $I=\{i_1, i_2, ..., i_M\}$ represent the set of users and items, where $N$ and $M$ represent the numbers of users and items. In the following, we will offer detailed definitions and concepts of KAMCL.

\textbf{Multi-Behavior Interaction Graph.} We denote the graph that illustrates the multi-behavioral interactions between user $u$ and item $i$ as $G_{ui}$. In more detail, $G_{ui} = \{U,\Theta,I\}$, where $\Theta=\{\theta^1,\theta^2,...\theta^B|\theta^b \in \mathbb{R}^{N \times M} \}$ represents the set of edges under different behaviors $b$ in $G_{ui}$, and $B$ is the number of behaviors. For example, when user $u$ interacts with item $i$ under behavior $b$, then $\theta_{u,i}^{b}=1$, and $\theta_{u,i}^{b}=0$ otherwise.

\textbf{Knowledge Graph.} The exceptional capabilities of knowledge graphs in supporting recommendations are harnessed as a fundamental tool in our study. This utilization aims to enrich item information and facilitate the process of intent generation. Specifically, we denote the knowledge graph by $G_{kg}=\{(h,r,t)|h,t \in E; r\in R\}$, where $E$ is the entity information corresponding to the item on the knowledge graph, $R$ is the set of relations, and $I\in E$. In the knowledge graph $G_{kg}$, each triple $(h,r,t)$ represents a semantic relation between a head-entity and a tail-entity. For instance, a triple $(Computer, Brand, Lenovo)$ represents that the brand of the computer is $Lenovo$. By fully leveraging this information, we can more accurately analyze the underlying reasons for users' interactions with items and extract information about their intentions.

\textbf{Task Formulations.} Based on the above definitions, we formally describe our task as follows:

\textbf{Inputs:} the user-item multi-behavior interaction graph $G_{ui}=\{U,\Theta,I\}$ and the knowledge graph $G_{kg}=\{(h,r,t)|\\h,t \in E; r\in R\}$.

\textbf{Outputs:} the score $\hat{\mathbf{y}}_{ui}$ that user $u$ interacts with item $i$ under the target behavior.

\begin{figure*}[t]
\centering
\setlength{\abovecaptionskip}{0cm}
\includegraphics[width=1\linewidth]{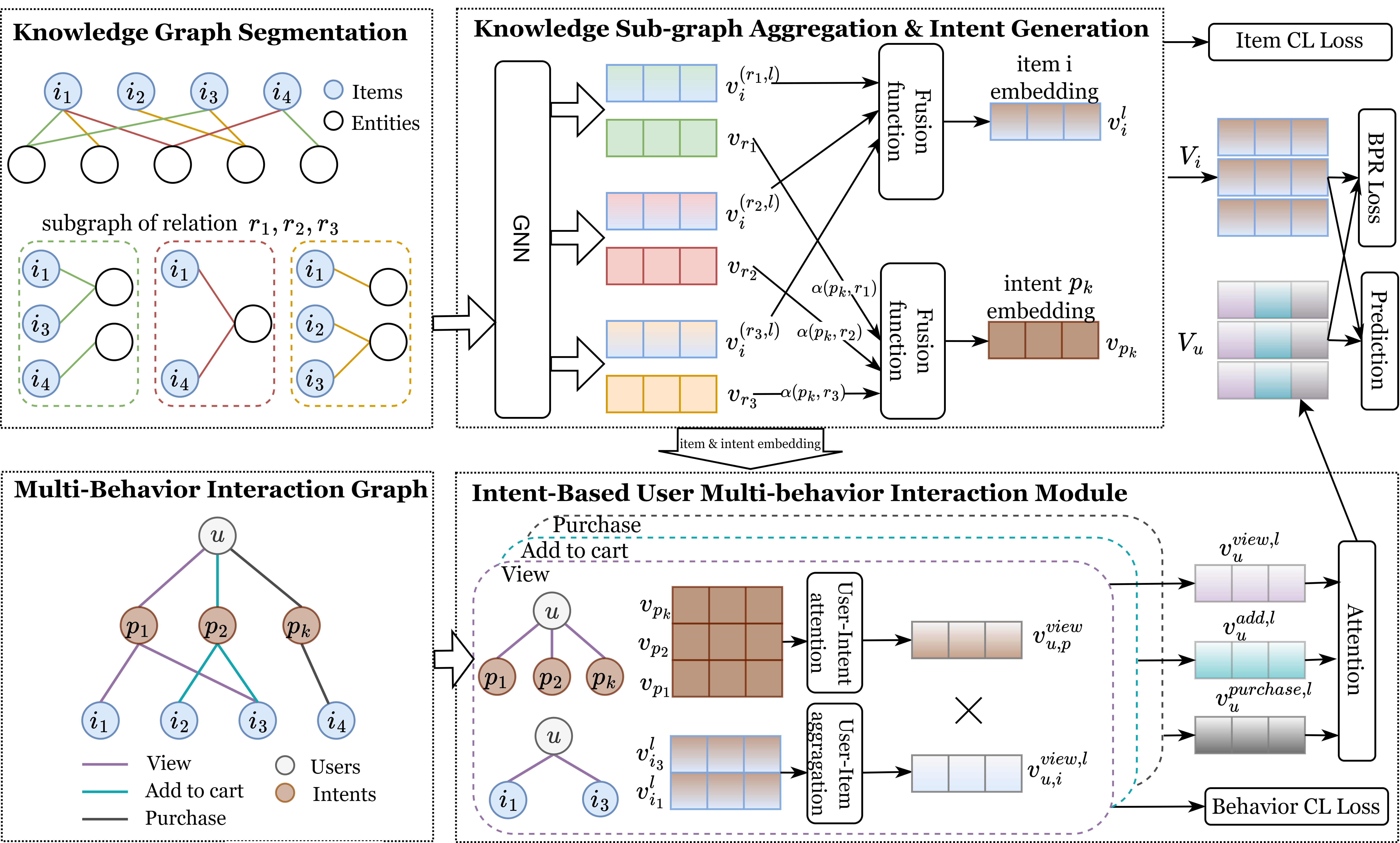}
\caption{The framework of KAMCL}\label{fig:model}
\label{figure}
\vspace{-0.8cm}
\end{figure*}

\section{METHODOLOGY}
In this section, we provide an overview of our model framework, depicting the model flow in Fig.~\ref{fig:model}. 

KAMCL consists of three modules: (1) relation-aware knowledge-graph aggregation module; (2) intent generate module; and (3) intent-based user multi-behavior interaction module. Specifically, both the relationship-aware knowledge graph aggregation module and the intent generation module take the knowledge graph as input to generate item embeddings and intent embeddings based on the entity and relationship information embedded therein, and then combine the user multi-behavioral interaction history to pass into the intent-based multi-behavioral interaction module to obtain the user's intent information under different behaviors to generate the recommended items under the target behaviors. Furthermore, we employ contrastive learning for joint training, enabling us to derive precise user and item representations and produce the final recommendation outcomes.

\subsection{Relation-aware Knowledge-graph Aggregation Module}
The knowledge graph plays a pivotal role in recommendation systems owing to its wealth of item-side information. Leveraging this feature, we've crafted a relation-aware knowledge graph aggregation module to glean semantic insights from diverse perspectives for items. The extracted outcomes are subsequently fused to acquire the ultimate item embedding.
\subsubsection{Relation-aware knowledge graph segmentation}
Effectively harnessing the rich semantic information within the knowledge graph to acquire item representations poses a significant challenge. Certain translation-based methods (e.g., TransE\cite{TransE}, TransR\cite{TransR}) represent entities and relations in a low-dimensional dense vector space, followed by computation and inference, extensively studied in recommender systems. Despite the proven effectiveness of these methods in practice, they come with drawbacks such as model complexity and limited interpretability. 

To address this, we meticulously partition the knowledge graph based on relation information. Specifically, we divide the knowledge graph into distinct sub-graphs according to their relations, and each sub-graph focuses on information from only one perspective (e.g., Brand) of the item:
\begin{equation}
\begin{aligned}
G_{kg}^{r_j} = \{(h, r_j, t) | h, t \in E; r_j \in R; j = 1,\dots,|R|\},\\ 
G_{kg} = \bigcup_{r \in R} {G_{kg}^r}.\ \ \ \ \ \ \ \ \ \ \ \ \ \ \ \ \ \ \ \ \ \ \ \ \ \ \ \ \ \ \
\end{aligned}
\end{equation}

where $|R|$ denotes the number of relationships in the knowledge graph. $r_j$ is a relation.

\subsubsection{Knowledge sub-graph propagation and aggregation}
After obtaining the knowledge graph for each relation, we apply graph neural network (GNN) technology to each sub-graph to extract information related to the characteristics of the item involved in each relation.

Specifically, we employ GNN to aggregate the neighborhood information of items in each sub-graph, enriching the representation of items under each relation with more comprehensive information. For each layer, we employ the following propagation method to update the representation of items:
\begin{equation}
\mathbf{v}_i^{(r_j, l)} = \frac{1}{|\mathcal{N}_i^{r_j}|} \sum\limits_{e \in \mathcal{N}_i^{r_j}} {\mathbf{v}_{e}^{(r_j, l-1)} \cdot \mathbf{v}_{r_j}}.
\end{equation}

where $\mathcal{N}_i^{r_j}$ is the set of first-order neighbors of item $i$ in knowledge sub-graph $G_{kg}^{r_j}$, and 
$\mathbf{v}_e^{(r_j,l-1)},\mathbf{v}_{r_j}\in\mathbb{R}^d$ denote the neighbor node embedding of item $i$ and relation embedding of $r_j$ in graph $G_{kg}^{r_j}$, and is also used as a projection operator for aggregating and propagating messages\cite{RotatE}. $d$ is the dimension of the embedding. $l$ is the $l$-th layer of GNN. Finally, after aggregating the propagated messages, we obtain the item embedding $\mathbf{v}_i^{(r_j, l)}$ under each relation.

As the knowledge sub-graphs under a single relation can only capture item information from one perspective, we fuse and transform the item embeddings obtained in each sub-graph. The specific formula is as follows:

\begin{equation}
\begin{aligned}
\mathbf{v}_{i}^{l} = fus(\mathbf{v}_{i}^{(r_1, l)} \| \mathbf{v}_{i}^{(r_2, l)} \| \dots \| \mathbf{v}_{i}^{(r_{|R|}, l)}), \\
fus(\mathbf{v}) = \mathbf{W}\mathbf{v} + \mathbf{bias}.\ \ \ \ \ \ \ \ \ \ \ \ 
\end{aligned}
\end{equation}

where $\|$ is the concatenation operation. $fus(\cdot)$ is a fusion function that connects the item embeddings in different relational sub-graphs. The projection matrix is denoted as $\mathbf{W} \in \mathbb{R}^{d \times |R|d}$, where $|R|$ is the number of relations. $\mathbf{v}\in\mathbb{R}^{|R|d}$ in this case denoting the item embedding obtained after the concatenation operation. 

\subsubsection{Relation-based contrastive optimisation}
Given the varying sparsity of item neighbor information in different knowledge sub-graphs, and considering that contrastive learning has demonstrated its effectiveness in addressing data sparsity issues\cite{wu2021self,zhu2021graph}. We propose a contrastive learning scheme to augment item representation. 

Due to differences in the information regarding item characteristics obtained from each knowledge sub-graph, we treat the item embedding acquired from different knowledge sub-graphs as a positive sample (denoted as $\{(\mathbf{v}_i^{(r,l)}, \mathbf{v}_i^{(r^{\prime},l)})|i\in I; r, r^{\prime} \in R, r \neq r^{\prime}\}$), while considering the embeddings of different items obtained from distinct sub-graphs as negative samples (denoted as $\{(\mathbf{v}_i^{(r,l)}, \mathbf{v}_{i^{\prime}}^{(r^{\prime},l)})|i, i^{\prime}\in I, i\neq i^{\prime}; r, r^{\prime} \in R, r \neq r^{\prime}\}$). To effectively address the sparsity differences in item representation, we employ the InfoNCE\cite{InfoNCE} loss function. The specific formula is as follows:
\begin{equation}
\mathcal{L}_{ICL} = \sum\limits_{i \in I}{\sum\limits_{\substack{{r,r^{\prime}}\in R,\\ \  r \neq r^{\prime}}} {-\log \frac{\exp (s(\mathbf{v}_i^{(r,l)}, \mathbf{v}_i^{(r^{\prime},l)})/\tau)}{\sum\limits_{\substack{i^{\prime} \in I,\\ i^{\prime}\neq i}}{\exp (s(\mathbf{v}_i^{(r,l)}, \mathbf{v}_{i^{\prime}}^{(r^{\prime},l)})/\tau)}}}}.
\end{equation}

where $s(\cdot)$ computes the similarity of two embeddings, and $\tau$ is a hyperparameter representing the temperature coefficient of InfoNCE.

\subsection{Intent Generate Module}
To harness information from the knowledge graph, we leverage the relationships within the graph to acquire intent representations. In this paper, we introduce an intent generation module that takes embedded relationships from the knowledge graph as input and produces corresponding intent embeddings.

\subsubsection{Intent representation learning}
Since different intents reflect the different concerns of the user about the item, for example, some intents prioritize the brand of the product, while others emphasize the price of the product. To better model this situation, inspired by KGIN\cite{KGIN2021}, we design intents as combinations of relations in a knowledge graph and use an attention mechanism to assist the generation of intents. $|P|$ is the number of intents, and the preliminary intent embedding $\mathbf{v}_{p_k}^{\prime} (p_k\in P, k=1,\dots,|P|)$ is computed as follows:
\begin{equation}
\begin{aligned}
\mathbf{v}_{p_k}^{\prime} = \alpha(p_k, r_1) \cdot \mathbf{v}_{r_1} \| \dots \| \alpha(p_k, r_{|R|}) \cdot \mathbf{v}_{r_{|R|}},\\
\alpha(p_k, r_j) = \frac{exp(w_{p_k, r_j})}{\sum\limits_{r^{\prime}\in R}exp(w_{p_k, r^{\prime}})}, \ \ \ \ \ \ \ \ \ \ \ \ \ \  
\end{aligned}
\end{equation}

where $\alpha(p_k,r_j)$ represents the attention coefficient of the relation $r_j$ in the intention $p_k$, which is computed from trainable weights $w_{p_k,r_j}$. In order to better explain the user interaction behavior, we use the same connection scheme as for the item. Finally, we join these relational embeddings to get the preliminary intent embedding $\mathbf{v}_{p_k}^{\prime} \in \mathbb{R}^{|R|d}$.

As the item representations are trained from knowledge sub-graphs of different relationships and intents are connected through these relationships, to enhance their alignment, we apply the same fusion layer as used in the processing of item embeddings to integrate the intent embeddings obtained above.
\begin{equation}
\mathbf{v}_{p_k} = fus(\mathbf{v}_{p_k}^{\prime}).  
\end{equation}

where $\mathbf{v}_{p_k}\in \mathbb{R}^d$ is the final obtained intent embedding.

\subsection{Intent-Based User Multi-behavior Interaction Module}
Users' intentions tend to be different under different behaviors. For example, in the e-commerce sector, users pay more attention to information such as brands and positive ratings when browsing for products, while they care more about the price when making a purchase. In order to better explore the connection between behaviors, we design an intent-based user multi-behavior interaction module. From the perspective of intent, we explore the connections and differences under different user behaviors.
\subsubsection{User Intent Generate for Different behaviors}
User interactions with items under different behaviors are motivated by distinct intentions. Consequently, the challenge lies in effectively capturing users' intentions across various behaviors. To tackle this issue, we employ the attention mechanism approach. The specific formula is as follows:
\begin{equation}
\begin{aligned}
\mathbf{v}_{u,p}^{b} = \sum\limits_{p_k\in P}{\beta(u,p_k) \cdot \mathbf{v}_{p_k}},\ \ \\
\beta(u,p_k) = \frac{exp(w^{b}_{u,p_k})}{\sum\limits_{p_k^{\prime}\in P}exp(w^{b}_{u,p_k^{\prime}})}.
\end{aligned}
\end{equation}

where $\beta(u,p_k)$ denotes the extent to which user $u$ is impacted by intention $p_k$ in behavior $b$, such as $view$, calculated from the trainable weights $w^{b}_{u,p_k}$. Ultimately, for each user $u$, we can combine the attention coefficient $\beta(u,p_k)$ and intent $\mathbf{v}_{p_k}$ to compute the user intent embedding $\mathbf{v}_{u,p}^{b}$ under each behavior $b$.

\subsubsection{Multi-behavioral interaction graph propagation and aggregation}
The items that users interact with under different behaviors reflect different preference information. To obtain this information, we use neighbor aggregation to generate corresponding user embeddings. The specific formula is as follows:

\begin{equation}
\mathbf{v}_{u,i}^{(b,l)} = \frac{1}{|\mathcal{N}_u^{b}|} \sum\limits_{i\in \mathcal{N}_u^{b}}{\mathbf{v}_i^l},
\end{equation}

where $\mathcal{N}_u^{b}$ denote the set of neighbors of user $u$ in the multi-behavioral interaction sub-graph $G_{ui}^{b} = \{U, \theta^{b}, I\}$, and $\mathbf{v}_i^l$ is the item embedding generated from the relation-aware knowledge graph aggregation module.

Then, we use the information acquired from user $u$ under behavior $b$ to fuse with the intent to obtain the user embedding.
\begin{equation}
\mathbf{v}_{u}^{(b,l)} = \mathbf{v}_{u,p}^{b} \cdot \mathbf{v}_{u,i}^{(b,l)} + \mathbf{v}_{u}^{(b,l-1)},
\end{equation}

According to the above formula, we can calculate the user embedding under each behavior. Then, we fuse these embeddings through an attention layer to get the final user embedding. The formula is as follows:

\begin{equation}
\mathbf{v}_{u}^{l} = \sum\limits_{b\in B}{\gamma_b \cdot \mathbf{v}_u^{(b,l)}}.
\end{equation}

where $\gamma_b$ is a hyper-parameter representing the weights of the different behavioral embeddings and $\mathbf{v}_{u}^{l} \in \mathbb{R}^d$ is the final user embedding.
\subsubsection{Behavior-based contrastive optimisation}
In the context of multi-behavior recommendation, while the target behavior may be sparse, it effectively reflects user preferences. Auxiliary behavior data, on the other hand, is abundant but comes with a high level of interference. Therefore, the key challenge lies in extracting meaningful preference information from these behaviors. An effective approach to address this issue is to employ a contrastive learning method to balance the differences between behaviors and enhance the performance of the target behavior.

In particular, we take the embeddings obtained under different behaviors of the same user as positive samples, denoted as $(\mathbf{v}_u^{(b,l)},\mathbf{v}_u^{(b^{\prime},l)}|u\in U; b,b^{\prime}\in B, b\neq b^{\prime})$, and the embeddings under different behaviors of different users as negative samples, denoted as $(\mathbf{v}_u^{(b,l)},\mathbf{v}_{u^{\prime}}^{(b^{\prime},l)}|u,u^{\prime}\in U, u\neq u^{\prime}; b,b^{\prime}\in B, b\neq b^{\prime})$. We use InfoNCE\cite{InfoNCE} as the loss function to construct a behavioral comparison-based learning scheme, which is calculated as follows:
\begin{equation}
\mathcal{L}_{BCL} = \sum\limits_{u \in U}{\sum\limits_{\substack{{b,b^{\prime}}\in B,\\ \  b \neq b^{\prime}}} {-\log \frac{\exp (s(\mathbf{v}_u^{(b,l)}, \mathbf{v}_u^{(b^{\prime},l)})/\tau)}{\sum\limits_{\substack{u^{\prime} \in U,\\ u^{\prime}\neq u}}{\exp (s(\mathbf{v}_u^{(b,l)}, \mathbf{v}_{u^{\prime}}^{(b^{\prime},l)})/\tau)}}}}.
\end{equation}

\subsection{Model Training and Prediction}
\subsubsection{Model prediction}
After the aggregation of the L-layer graph neural network, we can obtain the item and user representations of the different layers from the knowledge graph and user-item interaction graph. Considering that the user and item embeddings obtained from different network layers contain different embedding information, we will fuse the embeddings output from each layer to get the final user and item embeddings.
\begin{equation}
\mathbf{v}_u = \sum\limits_{l=1}^{L}{\mathbf{v}_u^l}, \mathbf{v}_i = \sum\limits_{l=1}^{L}{\mathbf{v}_i^l},
\end{equation}

where $\mathbf{v}_u$ and $\mathbf{v}_i$ denote the final user and item embedding, respectively, and $L$ denotes the number of layers in GNN. Then we calculate the score $\hat{\mathbf{y}}_{ui}$ of user $u$ interacting with item $i$ under the target behavior.
\begin{equation}
\hat{\mathbf{y}}_{ui} = \mathbf{v}_u^\top\mathbf{v}_i.
\end{equation}
\subsubsection{Loss optimisation}
To improve the recommendation accuracy of the model, we use BPR loss\cite{rendle2012bpr} for optimization. Specifically, under the target behavior, historical items that the user interacts with should have higher scores than those that do not.
\begin{equation}
\mathcal{L}_{BPR} = -\sum\limits_{(u,i)\in S^+}\sum\limits_{(u,i^{\prime})\in S^-}\log\sigma(\hat{\mathbf{y}}_{ui} - \hat{\mathbf{y}}_{ui^{\prime}}),
\end{equation}

where $(u,i)\in S^+$ is the set of positive samples with historical interactions under the target behavior and $(u,i^{\prime})\in S^-$ is the set of randomly generated negative samples. $\sigma(\cdot)$ is the sigmoid function.

Finally, combining the loss function constructed above with the two contrasting loss functions, we define the total loss function of the model as:
\begin{equation}
\mathcal{L} = \mathcal{L}_{BPR} + \lambda_1\mathcal{L}_{ICL} + \lambda_2\mathcal{L}_{BCL} + \lambda_3\|\Theta\|^2
\end{equation}

where $\lambda_1,\lambda_2$ are the hyper-parameters that control each loss. $\lambda_3$ is the coefficient of regularity term. ${\|\Theta\|^2}$ is the regularity term that prevents the model from overfitting.

\section{EXPERIMENT}
In this section, we conduct extensive experiments on three real-world datasets to answer the
following research questions:
\begin{itemize}
\item[$\bullet$]\textbf{RQ1}: How does KAMCL perform compared to other competitive baselines?

\item[$\bullet$]\textbf{RQ2}: How do different components of KAMCL affect its performance?

\item[$\bullet$]\textbf{RQ3}: How do hyper-parameters affect KAMCL?

\end{itemize}
\subsection{Experimental Settings}
\subsubsection{Dataset Description}

\begin{table}[t]
\footnotesize
\centering
\caption{Statistics of datasets}\label{tab:data_statistic}
\resizebox{\linewidth}{!}{
  \begin{tabular}{c|c|ccc}
    \toprule
    \multicolumn{2}{c|}{Dataset} & \multicolumn{1}{c}{Movielens} & \multicolumn{1}{c}{Yelp} & \multicolumn{1}{c}{Tmall} \\
    \midrule
    \midrule
    \multicolumn{1}{c|}{\multirow{4}[2]{*}{\makecell{User-Item\\Interaction}}} & \#User & 67,788 & 45,794 & 15,902 \\
          & \#Item & 8,704 & 43,653 & 53,721 \\
          & \#Interaction & $9.9\times10^6$ & $2.1\times10^6$ & $4.4\times10^5$ \\
          & \#Target & 4,970,984 & 1,291,921 & 51,608 \\
    \midrule
    \midrule
    \multicolumn{1}{c|}{\multirow{3}[2]{*}{\makecell{Knowledge\\Graph}}} & \#Entity & 11,636 & 48,144 & 103,469 \\
          & \#Relation & 4 & 25 & 4 \\
          & \#Triplet & 76,879 & 703,285 & 800,864 \\
    \bottomrule
    \end{tabular}
}
\vspace{-0.05in}
\end{table}

In order to verify the validity of our model, we evaluate it on three datasets Movielens\footnote{https://grouplens.org/datasets/movielens/10m/}\cite{KHGT2021}, Yelp\footnote{https://www.yelp.com/dataset/download}\cite{KHGT2021} and Tmall\cite{KMCLR2023}. The specific description of the dataset is shown in Table~\ref{tab:data_statistic}.

\begin{sloppypar}
For Movielens and Yelp, we adopt the same behavioral segmentation as in KHGT\cite{KHGT2021}. Specifically, for Movielens, we use user ratings of items to classify behaviors as $\{dislike,neutral,like\}$ and consider $like$ as the target behavior. For Yelp, we classify behaviors as $\{dislike,neutral,like,tip\}$ and consider $like$ as the target behavior. To better match the real situation, we select the Tmall dataset, which has a behavior type of $\{page\ view,favorite,add\ to\ cart,purchase\}$ and consider $purchase$ as the target behavior.
\end{sloppypar}

In this paper, since the knowledge graph is rich in item-side information, we use it to enrich the item representation. For the Movielens and Yelp datasets, following the processing in KB4Rec\cite{zhao2019kb4rec}, we use Freebase\footnote{https://developers.google.com/freebase/data} as knowledge graph. Additionally, for the Tmall dataset, we utilize the knowledge graph provided by KMCLR~\cite{KMCLR2023}. For these knowledge graphs and datasets, we adopt a standardized processing approach. Specifically, following the dataset filtering method in KHGT\cite{KHGT2021}, we iteratively filter entities and relations in the knowledge graph. Ultimately, we retain triples with a substantial number of relations to obtain the final dataset.

\subsubsection{Evaluation metrics}
\begin{sloppypar}
We employ Hit Ratio ($HR@K$) and Normalized\ Discounted\ Cumulative
Gain ($NDCG@K$) as evaluation metrics\cite{MMCLR2022,CML2022}, where a higher score indicates better model recommendations. We use the same methodology to split the three datasets. For each dataset, one item from the user's interactions under the target behavior is randomly selected as the test set, while the rest are used for training. To ensure efficient and fair model evaluation, we adopt the strategy used by ~\cite{KHGT2021,KMCLR2023}: selecting each positive item instance along with 99 randomly sampled non-interactive items for every user.
\end{sloppypar}

\begin{table*}[t]

  \centering
  \caption{This table shows the results of all methods on Movielens, Yelp, Tmall datasets. The best performance is boldfaced; the runner-up is underlined; and ‘Improve’ shows improvements achieved by KAMCL relative to the runner-up. $H@K$ and $N@K$ ($K=5,10$) denote the evaluation indicators HR and NDCG that we used.}\label{tab:result}
  \resizebox{\linewidth}{!}{
\begin{tabular}{p{7.28em}lcccccccccccc}
\toprule

    \multirow{2.2}[1]{*}{Method} & \multicolumn{4}{p{16em}}{\centering{Movielens}} & \multicolumn{4}{p{16em}}{\centering{Yelp}} & \multicolumn{4}{p{16em}}{\centering{Tmall}} \\
\cmidrule(lr){2-5}\cmidrule(lr){6-9}\cmidrule(lr){10-13}   \multicolumn{1}{p{7.28em}}{} & \multicolumn{1}{p{2em}}{H@5} & \multicolumn{1}{p{2em}}{N@5} & \multicolumn{1}{p{2em}}{H@10} & \multicolumn{1}{p{2em}}{N@10} & \multicolumn{1}{p{2em}}{H@5} & \multicolumn{1}{p{2em}}{N@5} & \multicolumn{1}{p{2em}}{H@10} & \multicolumn{1}{p{2em}}{N@10} & \multicolumn{1}{p{2em}}{H@5} & \multicolumn{1}{p{2em}}{N@5} & \multicolumn{1}{p{2em}}{H@10} & \multicolumn{1}{p{2em}}{N@10} \\
    \midrule
    MATN & 0.5792 & 0.4105 & 0.7468 & 0.4647 & 0.6569 & 0.4693 & 0.8204 & 0.5224 & 0.2861 & 0.1976 & 0.4016 & 0.2349\\
    MBGCN & 0.6714 & 0.4899 & 0.8197 & 0.5381 & 0.6854 & 0.4737 & 0.8465 & 0.5264 & 0.1578 & 0.1253 & 0.216 & 0.1438\\
    MBGMN & 0.6731 & 0.4867 & 0.8256 & 0.5364 & 0.7312 & 0.5352 & 0.8780 & 0.5832 & 0.3673 & 0.2859 & 0.4665 & 0.3178\\
    GNMR & 0.6536 & 0.4724 & 0.8050 & 0.5216 & 0.7641 & 0.5703 & 0.8785 & 0.6032 & 0.2106 & 0.1492 & 0.2939 & 0.1759\\
    KHGT & 0.6901 & 0.5016 & 0.8323 & 0.5481 & 0.7587 & 0.566 & 0.8813 & 0.6061 & 0.1268 & 0.0913 & 0.1815 & 0.1089\\
    CML & 0.4982 & 0.3298 & 0.6766 & 0.3872 & 0.6216 & 0.4365 & 0.7895 & 0.4914 & 0.2171 & 0.1526 & 0.3174 & 0.1849\\
    MBRec & \underline{0.7195} & \underline{0.5334} & \underline{0.8544} & \underline{0.5775} & 0.7564 & 0.5644 & 0.8836 & 0.6057 & 0.2141 & 0.1571 & 0.2871 & 0.1805\\
    CKML & 0.6426 & 0.4513 & 0.8012 & 0.5028 & \underline{0.7726} & \underline{0.5839} & \underline{0.8910} & \underline{0.6173} & \underline{0.4108} & \underline{0.3169} & \underline{0.5101} & \underline{0.3489}\\
    \midrule
    KAMCL & \textbf{0.7392} & \textbf{0.5498} & \textbf{0.8701} & \textbf{0.5923} & \textbf{0.7919} & \textbf{0.5981} & \textbf{0.9058} & \textbf{0.6353} & \textbf{0.5608} & \textbf{0.4714} & \textbf{0.6516} & \textbf{0.5006}\\
    \midrule
    Improve & 2.73\% & 3.07\% & 1.83\% & 2.56\% & 2.49\% & 2.43\% & 1.66\% & 2.91\% & 36.51\% & 48.75\% & 27.73\% & 43.47\%\\
    \bottomrule
    \end{tabular}%
    }
  \label{tab:addlabel}%

\end{table*}%

\subsubsection{Baselines}
We compare KAMCL to several state-of-the-art baselines. (1) Multi-behavior recommendation, including MATN, MBGCN, MBGMN, GNMR, CML, MBRec. (2) Knowledge-aware multi-behavior recommendation, including KHGT, CKML.
\begin{itemize}
\item[$\bullet$]\textbf{MATN}~\cite{MATN2020}: It applies attentional mechanisms to multi-behavioral recommendations and incorporates memory units to explore relationships between multiple behaviors.

\item[$\bullet$]\textbf{MBGCN}~\cite{MBGCN2020}: It represents user-item interactions as a graph, utilizing graph propagation to obtain behavior-aware embeddings, and employs graph neural networks for multi-behavior recommendation.

\item[$\bullet$]\textbf{MBGMN}~\cite{MBGMN2021}: It combines meta-learning with graphs to design a graph meta-neural network model. This model is used to explore the impact of multi-behavioral models on recommender systems.

\item[$\bullet$]\textbf{GNMR}~\cite{GNMR2021}: It makes further optimization on graphs and proposes a graph neural multi-behavioral enhancement network by introducing the diverse behavioral patterns of users through embedding propagation across the user-item interaction graph.

\item[$\bullet$]\textbf{KHGT}~\cite{KHGT2021}: It constructs a knowledge-aware item-item relation graph, and combines it with temporal information to explore the multi-behavioral recommendation problem through graph attention networks.

\item[$\bullet$]\textbf{CML}~\cite{CML2022}: It proposes a multi-behavioral recommendation framework based on contrast learning, designing a contrast meta-network to encode the heterogeneity between different behaviors.

\item[$\bullet$]\textbf{MBRec}~\cite{MBRec2022}: It specifically considers different interaction patterns and potential cross-type behavioral interdependencies, proposing an interrelational encoder to explain complex relational structures.

\item[$\bullet$]\textbf{CKML}~\cite{CKML2023}: It designs two modules, coarse-grained interest extraction and fine-grained behavioral correlation, which together capture the connection between behaviors.

\end{itemize}

\subsubsection{ Parameter settings}
We implement KAMCL with Pytorch and the model is optimized using the Adam optimizer during the training phase. For all modules, we use the same settings\cite{chen2021graph,KMCLR2023}: we set the user and item embedding dimensions to 64 and set the batch size to 128. For all datasets, we set the learning rate to 0.001. And the number of aggregation layers is searched from $\{1,2,3,4\}$. For loss function, $\lambda_1$ and $\lambda_2$ are searched from $\{1e^{-3}, 5e^{-3},1e^{-2}, 5e^{-2}\}$. $\lambda_3$ is searched from $\{1e^{-3}, 5e^{-3},1e^{-2}, 5e^{-2}, 1e^{-1}, 5e^{-1}, 1\}$. For intent number, it is searched from $\{1,2,4,8\}$, which is used in KGIN\cite{KGIN2021}. Other hyperparameters of the baseline models are adjusted according to the ranges provided in their respective papers.

\subsection{Comparative Results and Analysis. (RQ1)}
In Table~\ref{tab:result}, we show the experimental results of KAMCL with all baselines on the three datasets and summarise the following findings.

Comparing all baselines, our KAMCL achieves optimal results on datasets of different types and sparsities. This can be attributed to two main aspects. (1) The knowledge graph is introduced as auxiliary information to explore the connection between multiple user behaviors from the perspective of intent. (2) The design of a suitable comparative learning scheme, which effectively balances the differences between item attribute information and behavioral information.

As can be seen from the experimental results, CKML outperforms other neural network-based baselines like MATN, MBGCN, MBGMN, GNMR on most datasets. This is mainly due to the fact that unlike the traditional use of graphs only to model user interaction behaviors, CKML mines the connection between behaviors from a multi-granularity perspective with the help of knowledge graphs. Meanwhile, KHGT achieves good results on Movielens and Yelp datasets, but it is not applicable to the Tmall dataset due to the limitation of the way of constructing knowledge graphs. From these two points, it can be seen that our KAMCL can effectively use the information from the knowledge graph to explore the correlation between behaviors from the perspective of intent to achieve better performance.

CML does not achieve excellent results on the Movielens, Yelp dataset, which may be due to the fact that a scheme that relies only on learning by comparison between behaviors is not effective in capturing heterogeneity between behaviors when there are large differences between them (e.g., dislike, like). Our KAMCL is optimized on the basis of CML by adding item comparison learning on top of its original comparison scheme, and the two assist each other to better achieve the purpose of self-supervised learning.

In summary, our KAMCL outperforms existing state-of-the-art models (e.g., MBRec, CKML) on all three datasets. In particular, the advantage of our model is more obvious on the data sparse Tmall dataset. This can reflect the excellent performance of our model in dealing with sparsity problems.

\begin{table}[!htbp]
\centering
\renewcommand{\arraystretch}{1.3}
\caption{Results of ablation experiments}\label{tab:ablation}
\resizebox{\linewidth}{!}{
\begin{tabular}{c|cc|cc|cc}
\hline
\multirow{2}*{\diagbox[width=8em]{model}{dataset}} & \multicolumn{2}{c|}{Movielens} & \multicolumn{2}{c|}{Yelp} & \multicolumn{2}{c}{Tmall} \\
\cline{2-3}\cline{4-5}\cline{6-7}   \multicolumn{1}{c|}{} & \multicolumn{1}{|c}{H@10} & \multicolumn{1}{c|}{N@10} & \multicolumn{1}{c}{H@10} & \multicolumn{1}{c|}{N@10} & \multicolumn{1}{c}{H@10} & \multicolumn{1}{c}{N@10} \\
\hline
    w/o-KCL & 0.8602 & 0.5864 &0.8958 & 0.6208 & 0.3705 & 0.2665 \\
    w/o-KAL & 0.8623 & 0.5890 & 0.8978 & 0.6267 & 0.3833 & 0.2809 \\
    w/o-CL & 0.8657 & 0.5910 & 0.9049 & 0.6321 & 0.6399 & 0.4845 \\
    \hline
    KAMCL & \textbf{0.8701} & \textbf{0.5923} & \textbf{0.9058} & \textbf{0.6353} & \textbf{0.6516} & \textbf{0.5006}\\
\hline
\end{tabular}
}
\end{table}

\subsection{Ablation Study (RQ2)}
To verify the effect of different components on performance, we design three KAMCL variants for ablation experiments:

\begin{itemize}
\item[$\bullet$]w/o-KCL: 
it removes the knowledge graph and comparative learning scheme. Specifically, we are similar to KHGT by replacing the original knowledge graph-based part with a knowledge-aware item-item relationship graph and removing the comparative learning scheme. This is mainly to validate the joint impact of knowledge graph and comparative learning on performance.
\item[$\bullet$]w/o-KAL:  
it is primarily designed to validate the impact of the knowledge graph on performance. Its treatment is similar to that of w/o-KCL in that the original model knowledge graph is partially replaced with a knowledge-aware item-item relationship graph. 
\item[$\bullet$]w/o-CL: 
it removes the contrast learning scheme we designed. The main role of this variant is to validate the effectiveness of the contrast learning scheme in the model and to observe whether it has a positive impact on the final results.

\end{itemize}

Table~\ref{tab:ablation} shows the results of the ablation experiments. From the experiments, it can be seen that on the sparse behavioral Tmall dataset, the results of the KAMCL and w/o-CL using the knowledge graph are significantly better than the unused ones. Hence, it can be seen that the information in the knowledge graph has a positive effect on improving recommendation accuracy, especially in the case of sparse dataset, and helps the model make accurate recommendations. Comparing w/o-KCL with w/o-KAL, and w/o-CL with KAMCL, it can be seen that models employing the contrast learning scheme consistently outperform those that are not used. This shows the effectiveness of our designed contrast learning scheme, which aids the model to produce superior results on all datasets.

\subsection{Hyperparameter Analysis (RQ3)}
We show the impacts of some important hyper-parameters, including the number of aggregation layers $L$, the embedding dimension $d$ and the number of intents $|P|$.

\textbf{Impact of aggreation layer $L$}. 
Table~\ref{tab:paramter_layer} shows the results for different aggregation layers $L$ on the three datasets. It can be found that the more the number of layers, the better the results. The reason for this phenomenon is that more aggregation layers mean more information is embedded in the item and user representation. However, too much information can lead to overfitting of the model, therefore we chose 3 as the final result.

\begin{table}[!htbp]
\centering
\small
\renewcommand{\arraystretch}{1.3}
\caption{Impact of the number of layer $L$}\label{tab:paramter_layer}
\resizebox{\linewidth}{!}{
\begin{tabular}{c|cc|cc|cc}
\hline
\multirow{2}[1]{*}{\diagbox[width=8em]{model}{dataset}} & \multicolumn{2}{c|}{Movielens} & \multicolumn{2}{c|}{Yelp} & \multicolumn{2}{c}{Tmall} \\
\cline{2-3}\cline{4-5}\cline{6-7}   \multicolumn{1}{c|}{} & \multicolumn{1}{c}{H@10} & \multicolumn{1}{c|}{N@10} & \multicolumn{1}{c}{H@10} & \multicolumn{1}{c|}{N@10} & \multicolumn{1}{c}{H@10} & \multicolumn{1}{c}{N@10} \\
\hline
    KAMCL-1 & 0.8470 & 0.5650 & 0.8870 & 0.5875 & 0.4955 & 0.3293 \\
    KAMCL-2 & 0.8616 & 0.5836 & 0.8972 & 0.6094 & 0.5995 & 0.4378 \\
    KAMCL-3 & 0.8701 & 0.5923 & 0.9058 & 0.6353 & 0.6516 & 0.5006 \\
    KAMCL-4 & 0.8646 & 0.5918 & 0.8901 & 0.6215 & 0.6442 & 0.4979 \\
\hline
\end{tabular}
}
\end{table}

\begin{figure}[t]
	\centering
	\footnotesize
\subfigure{
		\begin{tikzpicture}[xscale=0.28,yscale=0.28]
		\begin{axis}[
 	xlabel ={\fontsize{18}{14}\selectfont dimension $d$},
	    ylabel={\fontsize{18}{14}\selectfont H@10(\%)},
		xmin=1, xmax=4,
		ymin=86.0, ymax=87.5,
		xtick={1,2,3,4},
        xticklabels={16,32,64,128},
		yticklabel style={/pgf/number format/.cd,fixed,precision=3},
		ytick={86.3, 86.6 ,86.9, 87.2},
		ylabel style ={font = \Large},
		xlabel style ={font = \Large}, enlarge x limits= 0.08,
		scaled ticks=false,
		tick label style={font = \Large},
		legend style={at={(0.5,0.98)},font = \large, anchor=north,legend columns=4,draw=none},
		ymajorgrids=true,
		grid style=dashed,
		]
		\addplot[color=myblue,
		mark=square,
		line width=2pt]coordinates {
		  (1, 86.14)
            (2, 86.68)
            (3, 87.01)
            (4, 86.50)
		};
		\legend{\fontsize{18}{14}\selectfont Movielens}
		\end{axis}
		\end{tikzpicture}
	}
\subfigure{
		\begin{tikzpicture}[xscale=0.28,yscale=0.28]
		\begin{axis}[
 	xlabel ={\fontsize{18}{14}\selectfont dimension $d$},
	    ylabel={\fontsize{18}{14}\selectfont H@10(\%)},
		xmin=1, xmax=4,
		ymin=89.4, ymax=90.9,
		xtick={1,2,3,4},
        xticklabels={16,32,64,128},
		yticklabel style={/pgf/number format/.cd,fixed,precision=3},
		ytick={89.7, 90.0 ,90.3, 90.6},
		ylabel style ={font = \Large},
		xlabel style ={font = \Large}, enlarge x limits= 0.08,
		scaled ticks=false,
		tick label style={font = \Large},
		legend style={at={(0.5,0.98)},font = \large, anchor=north,legend columns=4,draw=none},
		ymajorgrids=true,
		grid style=dashed,
		]
		\addplot[color=myblue,
		mark=square,
		line width=2pt]coordinates {
		  (1, 89.77)
            (2, 90.43)
            (3, 90.58)
            (4, 90.33)
		};
		\legend{\fontsize{18}{14}\selectfont Yelp}
		\end{axis}
		\end{tikzpicture}
	}
\subfigure{
		\begin{tikzpicture}[xscale=0.28,yscale=0.28]
		\begin{axis}[
 	xlabel ={\fontsize{18}{14}\selectfont dimension $d$},
	    ylabel={\fontsize{18}{14}\selectfont H@10(\%)},
		xmin=1, xmax=4,
		ymin=55.0, ymax=68.0,
		xtick={1,2,3,4},
        xticklabels={16,32,64,128},
		yticklabel style={/pgf/number format/.cd,fixed,precision=3},
		ytick={58.0, 61.0 ,64.0},
		ylabel style ={font = \Large},
		xlabel style ={font = \Large}, enlarge x limits= 0.08,
		scaled ticks=false,
		tick label style={font = \Large},
		legend style={at={(0.5,0.98)},font = \large, anchor=north,legend columns=4,draw=none},
		ymajorgrids=true,
		grid style=dashed,
		]
		\addplot[color=myblue,
		mark=square,
		line width=2pt
		]coordinates {
		  (1, 56.67)
            (2, 62.47)
            (3, 65.16)
            (4, 63.77)
		};
		\legend{\fontsize{18}{14}\selectfont Tmall}
		\end{axis}
		\end{tikzpicture}
}

\subfigure{
		\begin{tikzpicture}[xscale=0.28,yscale=0.28]
		\begin{axis}[
 	xlabel ={\fontsize{18}{14}\selectfont dimension $d$},
	    ylabel={\fontsize{18}{14}\selectfont N@10(\%)},
		xmin=1, xmax=4,
		ymin=58.2, ymax=59.6,
		xtick={1,2,3,4},
        xticklabels={16,32,64,128},
		yticklabel style={/pgf/number format/.cd,fixed,precision=3},
		ytick={58.5, 58.8 ,59.1},
		ylabel style ={font = \Large},
		xlabel style ={font = \Large}, enlarge x limits= 0.08,
		scaled ticks=false,
		tick label style={font = \Large},
		legend style={at={(0.5,0.98)},font = \large, anchor=north,legend columns=4,draw=none},
		ymajorgrids=true,
		grid style=dashed,
		]
		\addplot[color=myblue,
		mark=square,
		line width=2pt]coordinates {
		  (1, 58.35)
            (2, 59.06)
            (3, 59.23)
            (4, 58.96)
		};
		\legend{\fontsize{18}{14}\selectfont Movielens}
		\end{axis}
		\end{tikzpicture}
	}
\subfigure{
		\begin{tikzpicture}[xscale=0.28,yscale=0.28]
		\begin{axis}[
 	xlabel ={\fontsize{18}{14}\selectfont dimension $d$},
	    ylabel={\fontsize{18}{14}\selectfont N@10(\%)},
		xmin=1, xmax=4,
		ymin=61.3, ymax=64.0,
		xtick={1,2,3,4},
        xticklabels={16,32,64,128},
		yticklabel style={/pgf/number format/.cd,fixed,precision=3},
		ytick={61.6, 62.2 ,62.8, 63.4},
		ylabel style ={font = \Large},
		xlabel style ={font = \Large}, enlarge x limits= 0.08,
		scaled ticks=false,
		tick label style={font = \Large},
		legend style={at={(0.5,0.98)},font = \large, anchor=north,legend columns=4,draw=none},
		ymajorgrids=true,
		grid style=dashed,
		]
		\addplot[color=myblue,
		mark=square,
		line width=2pt]coordinates {
		  (1, 61.62)
            (2, 62.77)
            (3, 63.53)
            (4, 63.37)
		};
		\legend{\fontsize{18}{14}\selectfont Yelp}
		\end{axis}
		\end{tikzpicture}
	}
\subfigure{
		\begin{tikzpicture}[xscale=0.28,yscale=0.28]
		\begin{axis}[
 	xlabel ={\fontsize{18}{14}\selectfont dimension $d$},
	    ylabel={\fontsize{18}{14}\selectfont N@10(\%)},
		xmin=1, xmax=4,
		ymin=35.0, ymax=55.0,
		xtick={1,2,3,4},
        xticklabels={16,32,64,128},
		yticklabel style={/pgf/number format/.cd,fixed,precision=3},
		ytick={40.0, 45.0 ,50.0},
		ylabel style ={font = \Large},
		xlabel style ={font = \Large}, enlarge x limits= 0.08,
		scaled ticks=false,
		tick label style={font = \Large},
		legend style={at={(0.5,0.98)},font = \large, anchor=north,legend columns=4,draw=none},
		ymajorgrids=true,
		grid style=dashed,
		]
		\addplot[color=myblue,
		mark=square,
		line width=2pt
		]coordinates {
		  (1, 38.43)
            (2, 47.12)
            (3, 50.06)
            (4, 48.96)
		};
		\legend{\fontsize{18}{14}\selectfont Tmall}
		\end{axis}
		\end{tikzpicture}
}
	\caption{Impact of dimension $d$}
        \label{fig:parameter_dim}
 \vspace{-0.4cm}
\end{figure}
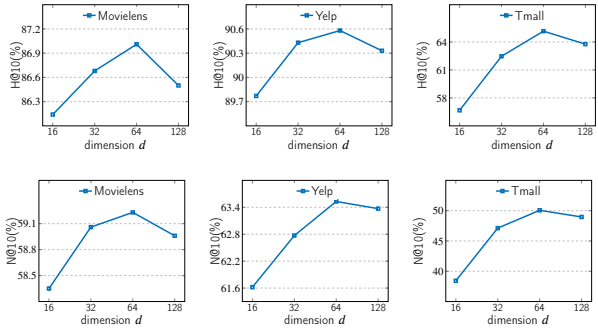

\textbf{Impact of embedding dimension $d$}.
 $d$ is searched from $\{16,32,64,128\}$. As can be seen from Fig.~\ref{fig:parameter_dim}, the results of $HR@10$ and $NDCG@10$ on the three datasets rise as the dimension $d$ increases. Larger dimension $d$ means that more information can be embedded, which is the same as what we expected. Also similar to the case of $L$, too much information can lead to a less effective model.

\begin{figure}[t]
	\centering
	\footnotesize
\subfigure{
		\begin{tikzpicture}[xscale=0.45,yscale=0.45]
		\begin{axis}[
 	xlabel ={intent number},
	    ylabel={H@10 Decrease(\%)},
		xmin=1, xmax=4,
		ymin=-1.8, ymax=0.5,
		xtick={1,2,3,4},
        xticklabels={1, 2, 4, 8},
		yticklabel style={/pgf/number format/.cd,fixed,precision=3},
		ytick={-1.2, -0.6, 0},
		ylabel style ={font = \Large},
		xlabel style ={font = \Large}, enlarge x limits= 0.08,
		scaled ticks=false,
		tick label style={font = \Large},
		legend style={at={(0.5,0.98)},font = \large, anchor=north,legend columns=4,draw=none},
		ymajorgrids=true,
		grid style=dashed,
		]
		\addplot[color=myblue,
		mark=triangle,
		line width=2pt]
        coordinates {
		  ( 1 , -0.04 )
            ( 2 , -0.02 )
            ( 3 , 0 )
            (4, -0.3)
		};
		\addplot[color=myyellow,
		mark=diamond,
		line width=2pt] coordinates {
    		( 1 , -1.71 )
            ( 2 , -0.71 )
            ( 3 , 0)
            ( 4 , -0.27)
		};
			\addplot[color=myred,
		mark=square*,
		line width=2pt] coordinates {
		  ( 1 , -0.89 )
            ( 2 , 0)
            ( 3 , -0.39)
            ( 4, -0.69)
		};	
\legend{Movielens, Yelp, Tmall}
		\end{axis}
		\end{tikzpicture}
	}
\subfigure{
		\begin{tikzpicture}[xscale=0.45,yscale=0.45]
		\begin{axis}[
 	xlabel ={intent number},
	    ylabel={N@10 Decrease(\%)},
		xmin=1, xmax=4,
		ymin=-2.4, ymax=0.5,
		xtick={1,2,3,4},
        xticklabels={1, 2, 4, 8},
		yticklabel style={/pgf/number format/.cd,fixed,precision=3},
		ytick={-1.8, -1.2, -0.6, 0},
		ylabel style ={font = \Large},
		xlabel style ={font = \Large}, enlarge x limits= 0.08,
		scaled ticks=false,
		tick label style={font = \Large},
		legend style={at={(0.5,0.98)},font = \large, anchor=north,legend columns=4,draw=none},
		ymajorgrids=true,
		grid style=dashed,
		]
		\addplot[color=myblue,
		mark=triangle,
		line width=2pt]
        coordinates {
		  ( 1 , -0.03 )
            ( 2 , -0.32 )
            ( 3 , 0 )
            (4, -0.32)
		};
		\addplot[color=myyellow,
		mark=diamond,
		line width=2pt] coordinates {
    		( 1 , -2.37)
            ( 2 , -2.31 )
            ( 3 , 0)
            ( 4 , -1.08)
		};
			\addplot[color=myred,
		mark=square*,
		line width=2pt] coordinates {
		  ( 1 , -0.89 )
            ( 2 , 0)
            ( 3 , -0.39)
            ( 4, -0.69)
		};	
\legend{Movielens, Yelp, Tmall}
		\end{axis}
		\end{tikzpicture}
	}
 
	\caption{Impact of intent number $|P|$
	}\label{fig:parameter_intent}
 \vspace{-0.8cm}
\end{figure}
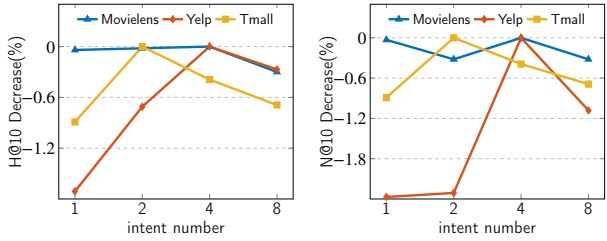

\textbf{Impact of intent number $|P|$}.
Similar to KGIN\cite{KGIN2021}, we conduct experiments on the effect of the number of intention on the results. Specifically, we select the values of $|P|$ from $\{1,2,4,8\}$ and conduct experiments on three datasets. The results of $HR@10$ and $NDCG@10$ are observed in Fig.~\ref{fig:parameter_intent}. On the Movielens dataset, the value of $|P|$ does not have a significant impact on the results, which may be due to the fact that the knowledge graph used in the Movielens dataset is too sparse. This makes the representation of intent and item not fit well. Whereas on the Yelp and Tmall datasets, it can be observed that an optimal number of intents leads to improved results. Excessively large values of $|P|$ elevate the similarity between intents, thereby causing a decline in model performance.

\section{CONCLUSION}

In this paper, we propose KAMCL for the multi-behavior recommendation. Our model employs knowledge graph information to establish connections between behaviors from an intent perspective. Specifically, our relationship-aware knowledge graph aggregation module obtains item representations from entity information, while the intent generation module acquires intent representations through relationship information. These representations are then combined with user multi-behavioral interaction data to reveal inter-behavioral connections at the granularity of intent. We develop a tailored comparative learning scheme to tackle sparsity issues in knowledge graphs and user multi-behavioral interaction data. Finally, we conduct experiments on three realistic datasets, demonstrating the superior performance of our model.

\bibliographystyle{cas-model2-names}

\bibliography{KAMCL}

\bio{}
\endbio

\bio{}
\endbio

\end{document}